\documentclass[a4paper,12pt]{article}
\pdfoutput=1 

\usepackage{jheppub} 
 \usepackage{tikz}
\usetikzlibrary{hobby}
\usepackage[T1]{fontenc} 
\usetikzlibrary{decorations.markings,intersections,patterns}
\usepackage{epsfig,amsfonts,amssymb,setspace}
\usepackage{tikz-cd}
\usepackage{relsize}

\title{ \centering  Quantum  Closed Superstring Field Theory and Hyperbolic Geometry I\\ {\large Construction of String Vertices}}

 \author{\centering Roji Pius}
\affiliation{Perimeter Institute for Theoretical Physics, Waterloo, ON N2L 2Y5, Canada}
\emailAdd{rpius@perimeterinstitute.ca}

\abstract{The complete quantum theory of closed superstrings is constructed using string diagrams endowed with metric having constant curvature $-1$.  The elementary string diagrams are equipped with the analytic local coordinates induced from the hyperbolic metric and  a distribution of a set of picture changing operators constructed using the identities satisfied by the simple closed geodesics on a hyperbolic Riemann surface. However, a slight modification near the boundary of the string vertices is needed to make them satisfy the  geometric condition imposed by the  quantum Batalin-Vilkovisky master equation.  }

\begin{document} 
\maketitle
\flushbottom

\section{Introduction}
\label{sec:intro}

The most attractive feature of conventional closed string perturbation theory is that it gives the total contribution of all $g$-loops graphs to any scattering amplitude with $n$ external states as a single integral over the space of inequivalent  Riemann surfaces with $n$ punctures and $g$ handles. However, it has  the following undesirable feature. In addition to the diagrams that are supposed to be included in scattering amplitude calculation, it also includes the contributions of the  diagrams, in which the radiative corrections are inserted into the external lines. Such diagrams have an internal line, connecting the self-energy part to the rest of the diagram. Due to the momentum conservation, the momentum flowing through this internal line is forced to satisfy the particle's mass shell condition, so that the propagator is evaluated at the mass-shell and provide infinite contribution to the amplitude \cite{Weinberg}. As a result, the amplitudes in conventional string perturbation theory are divergent in the presence of radiative corrections.\par

This undesirable feature of string theory can be cured, if it is possible to represent the string amplitude as a sum of separate terms. More precisely, if we are able to decompose the string amplitude into different pieces with each piece having a Feynman diagram interpretation, then we can throw away the contributions from the unwanted diagrams and obtain the correct scattering amplitude in string theory. The amplitudes defined this way are free from the infrared divergences. Interestingly, this is precisely what string field theory does: it decomposes the string amplitude into pieces, with each piece arising from a particular Feynman diagram of  string field theory.  In this sense, we can consider string field theory as the refined definition of string theory, formulated in the language of quantum field theory.  It generates the perturbative definition of string theory, starting from an action constructed using the gauge principle \cite{Witten:1985cc,9206084}. As we already discussed, since string field theory is formulated in the language of quantum field theory, following the standard techniques in quantum field theory such as the mass renormalization and the vacuum shift  \cite{Pius:2013sca,Pius:2014iaa,Pius:2014gza}, it can be used to compute the S-matrix elements of string theory, that are free from infrared divergences. The S-matrix elements computed this way with an appropriate  contour prescription for the loop momentum integration \cite{Pius:2016jsl,Pius:2018crk}, satisfy the unitarity requirement \cite{Sen:2016bwe,Sen:2016uzq}. Moreover, since string field theory is based on an action, it  can potentially pave the path towards the non-perturbative regime of string theory.  For instance, open string field theory has been successfully  used for studying the tachyon condensation phenomena in  open string theory \cite{Schnabl:2005gv}, though similar attempt in the case of closed string has not produced any definite result yet \cite{Yang:2005rx}. In principle, it may also be used to find backgrounds that are not acheivable from supergravity. \par

Covariant quantum field theories for both bosonic open  and closed strings have been constructed 
 a few decades ago, using the Batalin-Vilkovisky (BV) formalism \cite{Witten:1985cc,Thorn:1988hm,Bochicchio:1986zj,Saadi:1989tb,Kugo:1989aa,Kugo:1989tk,9206084,Hata:1993gf}. Recently, this construction has been successfully generalized to the case of heterotic and type II strings \cite{Sen:2015uaa}. There were  many attempts to achieve this generalization in  the past \cite{Witten:1986qs, Berkovits:1995ab,Berkovits:2001im,Okawa:2004ii,Berkovits:2004xh,Kroyter:2012ni,Matsunaga:2013mba,Erler:2013xta,Kunitomo:2013mqa,Erler:2014eba,Erler:2014eba,Matsunaga:2014wpa,Kunitomo:2014qla,Kunitomo:2014qla, Erler:2015rra,Erler:2015uba,Erler:2015lya,Goto:2015hpa,Kunitomo:2015usa}. The main stumbling block for the generalization was the following fact:  although, the Neveu-Schwarz (NS) sector of the closed superstrings/ heterotic superstrings allows the natural generalization of the construction in the bosonic case, the Ramond (R) sector forbids it by  leading us to an inconsistent kinetic term involving the Ramond sector fields. The construction in \cite{Sen:2015uaa} bypasses  this difficulty by introducing a set of auxiliary  fields in the Ramond sector, and then imposing a constraint on the external states that removes the extra states associated with the additional  fields \cite{Sen:2015hha}.  These auxiliary fields satisfy the free field equations, and hence once they are removed, they are not produced by the interactions, as required for the consistency. \par

The construction of closed superstring field theory described in \cite{Sen:2015uaa} was based on a set of crucial assumptions. Therefore, completing the construction of closed superstring field theory requires proving all these assumptions.  Before spelling out the assumptions, let us briefly explain the basic geometric ingredient of closed superstring field theory. The interaction terms in the quantum BV master action for closed superstring field theory are  defined by integrating the off-shell superstring measure over a set of inequivalent string diagrams, with the same number of  handles and punctures, describing the elementary interactions of the closed superstring states.  The collections of all such inequivalent string diagrams with a specific number of handles and a set of punctures form the {\it string vertices}. Every elementary string diagrams must be equipped with two sets of data. One set of data encodes the {\it choice of analytic local coordinates defined up to a constant phase around the punctures on each string diagrams}. The second set of data specifies the {\it distribution of the appropriate number of picture changing operators on each string diagrams}. The data corresponds to the choice of local coordinates must be independent of the labelling of the punctures, and the data corresponds to the distribution of picture changing operators must be invariant under the permutation of the NS punctures (punctures, where the states belong to the NS sector are inserted) and the R punctures (punctures, where the states belong to the R sector are inserted). Such a distribution of the picture changing operators must also be invariant under the action of an arbitrary large diffeomorphisms on the string diagrams. Moreover,  the physical quantities in the quantum closed superstring field theory  are gauge invariant, only if the action satisfies the quantum BV master equation.  This condition induces a very stringent {\it geometric condition upon the string vertices}, that are used for constructing the elementary vertices of closed superstring field theory.  {\bf The construction described in \cite{Sen:2015uaa} assumes the existence of such a choice of string vertices.} \par

 In principle, the string vertices in closed bosonic string field theory satisfying the  geometric identity imposed by the quantum BV master equation can be constructed by using the local coordinates induced from the metric of least possible area under the condition that lengths of all the nontrivial closed curves on the surface be greater than or equal to $2\pi$ \cite{9206084}. For Riemann surfaces with no handles, it is known that the minimal area metrics arise from the Jenkins-Strebel quadratic differentials \cite{Strebel}. Unfortunately, it is a daunting task to determine the Strebel differentials explicitly \cite{Moeller:2004yy}. In the case of Riemann surfaces with arbitrary number of handles, neither a convenient characterization nor a proof of existence of the minimal area metric is available yet \cite{Wolf:1992bk}.  \par

In order to perform computations in closed string field theory,  a clearly expressed description of the string diagrams characterizing the string vertices must be found. Furthermore, the  parametrization of the moduli space of the string diagrams  as well as the rules for performing integrations over this moduli space must be obtained.   In order to achieve this, there are two routes: either investigate more and develop the detailed theory of minimal area metrics {\footnote{Recently,  the convex optimization technique has been proposed for constructing the minimal area metric numerically for Riemann surfaces with handles \cite{Headrick:2018ncs,Headrick:2018dlw}.}}, or find an alternate construction of string vertices.  This paper follows the latter route. We describe an alternate construction of the string vertices using Riemann surfaces endowed with hyperbolic metric, metric having $-1$  curvature all over the surface \footnote{Another interesting approach to the construction of the string vertices is discussed in \cite{Erler:2017pgf}, where the cubic string vertex and one loop tadpole vertex for heterotic string field theory are constructed using the $SL\left(2,\mathbb{C}\right)$ local coordinate maps.}.\par 

The celebrated Riemann mapping theorem states that any simply connected open subset of the complex plane is biholomorphic to the upper-half plane \cite{Krantz}. This implies that  the local model of any  Riemann surface with negative Euler characteristic is the upper-half plane with the Poincar\'e-metric of constant curvature $-1$. In fact, the uniformization theorem states that all such Riemann surfaces are quotients of the upper-half plane \cite{Abikoff}. Therefore, a compact Riemann surface  having punctures and negative Euler characteristic has a complete hyperbolic metric, the metric with $-1$  curvature all over the surface \cite{Ahlfors1}. The most attractive feature of Riemann surfaces with hyperbolic metric is that, an explicit construction of such surfaces with arbitrary number of handles and punctures is readily available. They can be obtained by the proper discontinuous action of a Fuchsian group on the Poincar\'e upper half-plane \cite{FN,Taniguchi}. A Fuchsian group is a subgroup of the automorphism group of the Poincar\'e metric on the upper half-plane \cite{Maskit3}. Moreover, recently there is an outstanding progress in the theory of the Weil-Petersson symplectic geometry of the moduli space of the  Riemann surfaces with hyperbolic metric \cite{Wolpert1}, and it is proved to be very well suited for performing explicit integrations over the moduli space \cite{Mirzakhani1}.\par

Stimulated by these stupendous features of Riemann surfaces with hyperbolic metric,  we ask the following question: Is it possible to formulate quantum closed superstring field theory by considering string diagrams with hyperbolic metric on it? If  possible, then that would be a calculable realization of quantum closed superstring field theory \footnote{A formulation of bosonic string amplitudes in terms of hyperbolic geometry and Fenchel-Nielsen coordinates was attempted a long time ago in \cite{Eric1,Eric2} }. It was shown in \cite{Moosavian:2017qsp, Moosavian:2017fta, Moosavian:2017sev} that, it is indeed possible to formulate quantum closed bosonic string field theory in terms of string diagrams with hyperbolic metric on it. However, the local coordinates around the punctures induced from the $-1$ constant curvature metric (hyperbolic metric) defined on the string diagrams,  has to be  slight modified for those string diagrams that belongs to the boundary of the string vertices, in order to make them satisfy the geometric condition, imposed by the BV equation. \par 

In this paper, we generalize the construction  in \cite{Moosavian:2017qsp,  Moosavian:2017sev} to closed superstring field theory. The only additional geometric feature in the superstring case compared to the bosonic string case is the additional data regarding the distribution of picture changing operators on the elementary string diagrams.   We describe a systematic method for distributing picture changing operators consistently explore the beautiful identities satisfied by the simple closed geodesics on  Riemann surfaces with hyperbolic metric \cite{McShane1,Mirzakhani1} to obtain an MCG invariant  distribution of picture changing operators. Like in the case of local coordinates, it is necessary to  slightly modify this  distribution  of picture changing operators on string diagrams that belongs to the boundary regions of the string vertices, to make it satisfy the geometric condition arising from the BV master equation.   \par

This paper is organized as follows: In section \ref{BV}, we review the general construction of the quantum BV master action for closed superstring field theory. In section \ref{cons},  we describe the construction of string vertices in closed superstring field theory using Riemann surfaces endowed with metric having constant curvature $-1$.

\section{BV Master Action for Closed Superstring Field Theory}\label{BV}

In this section, we  briefly review the construction quantum BV master Action for the covariant string field theories for heterotic and type II superstring theories using of the picture changing formalism for RNS-superstring theory \cite{Sen:2015uaa}. 

\subsection{World-sheet superconformal field theory}

The superstring field theory is formulated using a superconformal field theory (SCFT) defined on a Riemann surface. In order to avoid cluttering we only discuss the construction of  closed heterotic string field theory. The generalization to Type II superstring theories is straightforward. The world-sheet SCFT of the heterotic string theory consist of a matter sector, given by the SCFT with the central charge $(26,15)$ and a ghost sector, given by the SCFT with the central charge $(-26,-15)$. The right-moving  stress tensor of the matter sector is denoted by $T_m(z)$ and its superpartner by $T_F(z)$. Similarly, the left moving stress-tensor of the matter sector is denoted by $\overline T_m(\bar z)$. The ghost system  contains the anti commuting fields $b(z),c(z),\bar b(\bar z),\bar c(\bar z)$ and the commuting $\beta(z),\gamma(z)$ ghosts. The stress tensors of the ghost fields are given by 
\begin{align}\label{betagbc stress tensors}
T_{b,c}(z)&=-2b(z)\partial c(z)+c(z)\partial b(z) \nonumber\\ \overline T_{\bar b,\bar c}(\bar z)&=-2\bar b(\bar z)\bar \partial \bar c(\bar z)+\bar c(\bar z)\bar \partial \bar b(\bar z)\nonumber\\
T_{\beta,\gamma}(z)&=\frac{3}{2}\beta(z)\gamma(z)+\frac{1}{2}\gamma(z)\partial \beta(z)
\end{align}
 
 The ghost fields $c(z)$ and $\bar c(\bar z)$  have the conformal dimensions  $(-1,0)$ and $(0,-1)$ respectively and the anti-ghost fields $b(z)$ and $\bar b(\bar z)$ have the conformal dimensions $(2,0)$ and $(0,2)$ respectively. They have the following  mode expansions:
  \begin{align}\label{modebc}
  c(z)&=\sum_n\frac{c_n}{z^{n-1}} \qquad\qquad \bar c(\bar z)=\sum_n\frac{\bar c_n}{\bar z^{n-1}}\nonumber\\
    b(z)&=\sum_n\frac{b_n}{z^{n+2}} \qquad \qquad\bar b(\bar z)=\sum_n\frac{\bar b_n}{\bar z^{n+2}}
  \end{align}
The non-vanishing anti-commutation relations satisfied by these modes are as follows:
  \begin{equation}\label{anticommbc}
  \{b_n,c_m\}=\{\bar b_n,\bar c_m\}=\delta_{m+n,0}
  \end{equation}

 The world-sheet SCFT has the following BRST operator
 \begin{equation}\label{BRSTB}
 Q_B=\oint \frac{dz}{2\pi\mathrm{i}}j_{B}(z)+\oint \frac{d\bar z}{2\pi\mathrm{i}}\bar j_{B}(\bar z)
 \end{equation} 
where the holomorphic BRST current $j_B(z)$ and the anti-holomorphic BRST current $\bar j_{B}(\bar z)$  given by
\begin{align}
j_B(z)&= c(z)\left(T_m(z)+T_{\beta,\gamma}(z)\right)+\gamma(z)T_F(z)+b(z)c(z)\partial c(z)-\frac{1}{4}\gamma(z)^2b(z)\nonumber\\
\bar j_B(\bar z)&=\bar c(\bar z)\bar T_m(\bar z)+\bar b(\bar z)\bar c(\bar z)\bar \partial c(\bar z)
\end{align}
Using the operator product expansion 
   \begin{equation}
   b(z)c(w)\sim \frac{1}{z-w}
   \end{equation}
we can verify that
   \begin{equation}\label{QbB}
   \{Q_B,b(z)\}=T(z) \qquad \qquad \{Q_B,\bar b(\bar z)\}=\overline T(\bar z)
   \end{equation}
   where $T(z)$ denotes the total stress-energy tensor for the world-sheet SCFT. This equations implies that 
   \begin{equation}\label{Qfirstc}
 Q_B=c_0^+L_0^++\bar c_0^+\overline L_0^++\cdots
 \end{equation}  
  The dots indicate terms that do not involve zero modes of the ghost fields and 
     \begin{equation}
L_0^{\pm}=L_0\pm\overline{ L}_0
   \end{equation}
   The operators $L_n$ and $\overline L_n$ are the total Virasoro generators in the left and right-moving sectors of the world-sheet theory.  The Virasoro generators are the modes of the following mode expansion of the total stress-energy tensor: 
   \begin{equation}
   T(z)=\sum_{n}\frac{L_n}{z^{n+2}} \qquad \qquad   \overline{T}(\overline{z})=\sum_{n}\frac{\overline{L}_n}{\overline{z}^{n+2}}
   \end{equation}

\subsection{Concept of pictures}\label{picture}
The intricate nature of the picture changing formulation of the superstring theory stems from the curious properties of the $\beta\gamma$ system, and hence it is appropriate to discuss the $\beta\gamma$ system and its representations in detail. The $\beta\gamma$ system  is a commuting fermionic system with  mode expansion for the fields  given by
\begin{align}\label{beta}
\beta(z)&=\sum_n \beta_n z^{-n-\frac{3}{2}} \nonumber\\
 \gamma(z)&=\sum_n \gamma_n z^{-n+\frac{1}{2}} 
\end{align}
where $n$ is a half integer for the NS sector and an integer for the R sector. These modes satisfy the following commutation relations
\begin{equation}\label{betac}
[\gamma_m,\beta_n]=\delta_{n,-m}
\end{equation}
This algebra has  infinite number of inequivalent representations, and all these inequivalent representations can be constructed using the raising operators by acting on infinite number of vacuum states $|q\rangle$, where $q$ being the ghost charge of the vacuum states $|q\rangle$ is integer or half integer. The ghost charge $q\in \mathbb{Z}$ for the NS sector and $q\in \mathbb{Z}+\frac{1}{2}$ for the R sector, are  the eigenvalue of the ghost charge operator given by
\begin{equation}
Q_{gh}=\sum_n \beta_n\gamma_{-n}
\end{equation}
Unlike in the case of  the degenerate ground states of the $bc$ system, here we can not go from the vacuum with one value of $q$ to another vacuum with a different value for $q$ by acting with finite number of oscillators, and hence they are inequivalent. \par

Let us denote the operators which can be used to increase or decrease the ghost charge of the vacuum $q$ by $\delta(\gamma_m)$ and $\delta(\beta_n)$. Their action on the $q$-vacua are given by
\begin{align}\label{rlqvacuum}
\delta(\beta_{-q-\frac{3}{2}})|q\rangle &= |q+1\rangle \nonumber\\
 \delta(\gamma_{q+\frac{1}{2}})|q\rangle &= |q-1\rangle
\end{align}
Similarly, we can define operators $\Sigma_+$ and $\Sigma_-$, the spin fields, for mapping  states in the R-sector  to the states in the NS-sector or vice versa.  They are also defined by their action on $q$-vacua given by
\begin{align}\label{spinf}
\Sigma_+(0)|q\rangle &= \left|q+\frac{1}{2}\right\rangle \nonumber\\
 \Sigma_-(0)|q\rangle &= \left|q-\frac{1}{2}\right\rangle 
\end{align}

Consequently, each of the states in the Hilbert space of superstring  theory has an infinite number of inequivalent representation based on the $q$-vacua that we use for building the tower of the states. From the operator-state correspondence, we know that there exist a vertex operator associated to the each state in the Hilbert space. As a result, there are infinite number of inequivalent vertex operators for any specific state and we discriminate each of them by associating a {\it picture number} which indicates the $q$-vacua used for constructing the states.  \par

The discussion is more transperant if we represent $\beta\gamma$ system using a free scalar $\phi$ and a pair of free chiral fermionic fields $\xi$ and $\eta$ of conformal weight (0,0) and (1,0), respectively. This changeover of the representation is known as the {\it bosonization} \cite{Friedan:1985ge}. The free scalar $\phi$ is compactified on the circle $R/2\pi Z$ and is coupled to a background charge $Q=2$. The action for the combined system is as follows
\begin{equation}\label{bgbos}
S[\phi,\xi,\eta]=\frac{1}{2\pi}\int \left( \partial \phi\bar\partial \phi-\frac{1}{2}R\phi\right)+\frac{1}{\pi}\int \eta\bar\partial\xi 
\end{equation} 
The stress-energy tensor is given by
\begin{equation}\label{betagbc stress tensors phi}
T_{\beta,\gamma}(z)=T_{\phi}+T_{\eta,\xi}
\end{equation}
where 
\begin{align}\label{stressetaxi}
T_{\eta,\xi}(z)&=-\eta(z)\partial \xi(z)\nonumber\\
 T_{\phi}(z)&=-\frac{1}{2}\partial \phi(z)\partial \phi(z)-\partial^2\phi(z)
\end{align}
The precise mapping between the two systems is as follows:
\begin{equation}\label{betatgammaboso}
\beta=e^{\phi}\partial_z\xi \qquad \gamma=e^{-\phi}\eta \qquad \delta(\beta)=e^{-\phi} \qquad \delta(\gamma)=e^{\phi} \qquad \xi=\mathrm{ H}(\beta) \qquad \eta=\partial_z\gamma\delta(\gamma)
\end{equation}
where $\mathrm{ H}$ denotes the Heaviside step function. Using these relations, it is not difficult to see that 
\begin{equation}\label{deltabeta}
\delta(\beta(z))=e^{\phi(z)} \qquad \delta(\gamma(z))=e^{-\phi(z)} \qquad \Sigma_+(z)=e^{\frac{1}{2}\phi(z)} \qquad \Sigma_-(z)=e^{-\frac{1}{2}\phi(z)}
\end{equation}
It is important to note that only the derivatives of $\xi$ field is present in the identification \ref{betatgammaboso} between the two systems. Therefore, the constant zero mode $\xi_0$ can not be produced from any of the operators acting within $\beta\gamma$ system. This means that the Hilbert space ${\cal H}_{\xi\eta\phi}$ of the $(\xi,\eta,\phi)$  is twice as large as the Hilbert space ${\cal H}_{\beta\gamma}$ of the $\beta\gamma$ system. The precise equivalence is the following
\begin{equation}\label{hilberteq}
{\cal H}_{\beta\gamma}=\left\{ |\psi\rangle\in {\cal H}_{\xi\eta\phi}~|~\eta_0|\psi\rangle=0\right\}
\end{equation} 
Since it is possible to construct the operators which convert one $q$-vacua to another, it should also be possible to change the ghost charge of a vertex operator. This procedure is known as the {\it picture-changing} operation.\par

 A general state in the Hilbert space built from the vacuum state with ghost charge $q$  can be converted to a state in the Hilbert space built from the vacuum state with ghost charge $q+1$ by acting with $G_{-q-\frac{3}{2}}\delta\left(\beta_{-q-\frac{3}{2}}\right)$ with $G_n$ defined by 
\begin{equation} 
T_F(z)=\frac{1}{2}\sum_n G_nz^{-n-\frac{3}{2}}
\end{equation}
On the physical vertex operators, i.e. those satisfy $\delta_{BRST}V^{(q)}=\partial(cV^{(q)})$, the {\it picture-changing operation} is represented as 
\begin{align}\label{picture}
V^{(q)}(w)\to V^{(q+1)}(w)&=\oint \frac{dz}{2i\pi }j_{B}\Big(\xi V^{(q)} \Big)(w)-\partial\Big( c\xi V^{(q)}\Big)(w)\nonumber\\
&=\chi(w)V^{(q)} (w)-\partial\Big( c\xi V^{(q)}\Big)(w)
\end{align}
where $V^{(q)}$ is the vertex operator with picture number $q$ and $j _{B}$ is the superstring BRST current. Although the second term in the right hand side looks like a BRST variation, it is not really a BRST exact deformation due to the presence of the zero mode of the $\xi$ field in it. Remember that the zero mode $\xi_0$ is not in the Hilbert space ${\cal H}_{\beta\gamma}$. The operator $\chi(z)$ is known as the {\it picture-changing operator} (PCO) and it is given by
\begin{align}\label{hpart}
\chi(z)=\{Q_B,\xi(z)\}=\oint dw~ j_{B}(w)\xi(z)
\end{align}
 The picture-changing operator is BRST invariant dimension zero primary operator having the picture number one. Before we end our discussion on pictures, let us note that it is necessary to introduce certain number of PCO's on the world-sheet to make sure that total picture number of the superstring measure is zero. The picture number anomaly demands that we must insert  \begin{equation}
 2g-2+\sum_{i=1}^{m_N}q^i_{N}+\sum_{j=1}^{m_R} q^j_{R}
 \end{equation}
 number of PCO's on a genus $g$ Riemann surface with $m_{N}$ number of NS punctures and $m_R$ number of  R punctures. Here,  $q^i_{N}$ is the picture number of the NS state inserted at the $i^{th}$ NS  puncture $q^j_{R}$ is the picture number of the R state inserted at the $j^{th}$ R  puncture.

\subsection{String fields }\label{dsupstringfield}

The basic degrees of freedoms in string field theory are the string fields. Let us denote the full Hilbert space of the world-sheet matter plus ghost SCFT carrying arbitrary picture and ghost numbers by $\mathcal{H}$.  An arbitrary string field is a vector in the Hilbert space $\mathcal{H}$ of the world-sheet SCFT.  The string fields that enter into the construction of the action for the  superstring field theory  are not completely arbitrary, they are required to satisfy following conditions  \cite{Sen:2015uaa,Sen:2016bwe}: 
\begin{itemize}
	\item Must be annihilated by $L_0^{-}=L_0+\bar L_0$ and $b_0^{-}=b_0+\bar b_0$
	
	\item Carry arbitrary ghost numbers
	
	\item Must have even Grassmannality
	
	\item Must satisfy an appropriate reality condition to ensure the reality of action.
	
	\item Must be in a specific picture 
	
\end{itemize}

 Now we shall state the specific reality condition that we choose to work with. Let us denote by $ |\Psi\rangle_{NS}$ and $ |\Psi\rangle_{R}$, the NS sector and R sector components of the string field $|\Psi\rangle$ :
	\begin{align}
	|\Psi\rangle_{NS}&= \sum_s|\phi_r\rangle\psi_r\nonumber\\
	|\Psi\rangle_{R}&=\sum_r|\hat\phi_s\rangle\hat\psi_s
	\end{align}
	The  string fields satisfy  the following reality condition:
	\begin{align}
	\psi_r(k)^*&=(-1)^{n_r(n_r+1)/2+1}\psi_r(-k)\nonumber\\
	\hat\psi_s(k)^*&=-\mathrm{i}(-1)^{(n_r+1)(n_r+2)/2}\hat\psi_s(-k)
	\end{align}
	where $n_r$ and $n_s$ denote the ghost numbers of $\phi_r$ and $\hat \phi_s$ respectively.\par
	
	Two kinds of string fields are required for constructing the action for closed superstring theory, the {\it dynamical string fields} $|\Psi\rangle$ and the {\it auxiliary string fields}  $|\tilde\Psi\rangle$. The auxiliary superstring fields enter only into the kinetic term of the action. The interaction terms are constructed only using the dynamical superstring fields.   The dynamical string field $|\Psi\rangle$ contains the NS states with picture number $-1$ and the R states with picture number $-\frac{1}{2}$. The auxiliary string field $|\widetilde\Psi\rangle$ contains the NS states with picture number $-1$ and the R states with picture number $-\frac{3}{2}$.\par

Let us explain the need for restricting the superstring fields to have only  $-1$ picture number in the NS sector and $-\frac{1}{2}$ and $-\frac{3}{2}$ picture numbers in the R sector \cite{deLacroix:2017lif}.  For the on-shell string states, the BRST cohomology is  the same in all picture numbers \cite{Friedan:1985ge}. Therefore, we can choose to work with any fixed picture number sector modulo certain ambiguities due to the boundary terms. The situation is slightly different in  superstring field theory. To appreciate the difference, consider the action  of the $\beta$ and $\gamma$ modes on the $q$ vacuum:
\begin{align}
\beta_n|q\rangle&=0 \qquad \mathrm{for}~n\geq -q-\frac{1}{2} \nonumber\\
 \gamma_n|q\rangle&=0 \qquad \mathrm{for}~n\geq q+\frac{3}{2}
\end{align}

This suggests that if $q=-1$ all of the positive modes of $\beta $  and $\gamma$ annihilate the vacuum. For any other integer picture number, there will be either some positive mode of $\beta$ or positive mode of $\gamma$ that will not annihilate the vacuum.  These modes can be used to create states of arbitrary negative conformal dimension. Remember that in  string field theory all off-shell states  propagate inside loops, including the states having arbitrary negative conformal dimension. This will make the theory inconsistent, and this can be avoided this by restricting the off-shell states in the NS sector to have only $-1$ picture number.  In the R sector the requirement of having a vacuum state that is annihilated by all the positive modes of  fields $\beta$ and $\gamma$ restrict the  off-shell states in the R sector to have picture numbers $-\frac{1}{2}$ and $-\frac{3}{2}$. Note that even after choosing  $-\frac{1}{2}$ or $-\frac{3}{2}$ picture numbers, the vacuum is not annihilated by the zero mode operators $\gamma_0$ and $\beta_0$. As a result, we can create infinite number of states at the same mass$^2$ level by applying $\gamma_0$ and $\beta_0$. Fortunately, this will not create unsurmountable difficulty if we take the interacting off-shell string states to have picture number $-\frac{1}{2}$. This is because the special form of the propagator for superstring field theory do not allow the propagation of all of these infinite number of states at the same mass$^2$ level \cite{deLacroix:2017lif}.

\subsection{String vertices }\label{sstrinvsi}

The interaction terms in the action for closed superstring field theory can be understood as  certain integrals over all possible distinct elementary string diagrams representing elementary string interactions. The most important geometric input needed for constructing the action for closed superstring field theory is the complete description of  the set of elementary string diagrams. The set of all distinct elementary string diagrams with fixed number of handles and  number of NS punctures and R punctures is called the {\it string vertex} for closed superstring theory. We denote such a string vertex by $\mathcal{W}_{g,m_N,m_R}$, where $g$ is the number of handles, $m_N$ is the number of NS punctures and $m_R$ is the number of R punctures. \par

An elementary string diagram that belongs to  $\mathcal{W}_{g,m_N,m_R}$ can be specified by  two set of data on a Riemann surface with $g$ handles and $m_N$ number of NS punctures and $m_R$ number of R punctures. One set of data specifies the local coordinates around the punctures on the Riemann surface.  The local coordinates around each punctures are only defined up to a constant phase. Up to this ambiguity, the local coordinates must be defined continuously over the set  $\mathcal{W}_{g,m_N,m_R}$. The other set of data specifies the locations on the Riemann surface at which the PCO's are placed.  Naively, the PCO's can be placed at any positions on a superstring world-sheet. A careful analysis reveals that the superstring measure can  have {\it spurious poles} \cite{Verlinde:1987sd}, when the vertex operators collide with the PCO's, the PCO's collide each other and the  locations of PCO's on the world-sheet satisfy certain global conditions. It is important to choose a distribution of PCO's on the elementary string diagram that does not give rise to any unphysical singularities.\par
 
String vertex $\mathcal{W}_{g,m_N,m_R}$ contains a set of Riemann surfaces that cover a connected region inside $\mathcal{M}_{g,m_N+m_R}$, the moduli space of Riemann surfaces with $g$ handles and $m_N+m_R$ number of punctures. This set does not include surfaces arbitrarily close to the degeneration.   The choice of superstring vertex $\mathcal{W}_{g,m_N,m_R}$ is not arbitrary. It must satisfy the following requirements:
\begin{itemize}
	\item The assignment of local coordinates around the punctures on the elementary string diagrams must be independent of the labeling of the punctures.  If a surface $\mathcal{R}$ with labeled punctures is in $ \mathcal{W}_{g,m_N,m_R}$ then copies of $\mathcal{R}$ with all other inequivalent  labelings of the punctures are also included in $ \mathcal{W}_{g,m_N,m_R}$;
	
	\item If $\mathcal{R}\in  \mathcal{W}_{g,m_N,m_R}$ then $\mathcal{R}^*\in \mathcal{W}_{g,m_N,m_R}$, where $\mathcal{R}^*$ denotes the mirror image of $\mathcal{R}$. The local coordinates in $\mathcal{R}$ and $\mathcal{R}^*$ are related by the antiholomorphic map (i.e. $z\longrightarrow -\bar{z}$) that relates the two surfaces;
	
	\item The distribution PCO's on each surface belongs to $ \mathcal{W}_{g,m_N,m_R}$  must avoid the occurrence of spurious poles;
	
	\item The distribution PCO's on each world-sheet belongs to $ \mathcal{W}_{g,m_N,m_R}$  must be invariant under large diffeomorphisms performed on the world-sheet.

\end{itemize}
It is possible to  construct string vertices avoiding the occurrence of  spurious poles systematically by using the vertical integration prescription  introduced in\cite{Sen:2014pia} and elaborated in \cite{Sen:2015hia}.\par~\par

\noindent{\bf{\underline{Geometric Identity}} :} The string vertices $\mathcal{W}$ can be used to construct the action satisfying BV master equation only if they satisfy the following very stringent {\it geometric identity}
	\begin{align}\label{bvmastercond1}
	\partial \mathcal{W}_{g,m,n}=&-\frac{1}{2}\mathop{\sum_{g_1,g_2}}_{g_1+g_2=g}\mathop{\sum_{m_1,m_2}}_{m_1+m_2=m+2}\mathop{\sum_{n_1,n_2}}_{n_1+n_2=n}\mathbf{ S}[\{ \mathcal{W}_{g_1,m_1,n_1}, \mathcal{W}_{g_2,m_2,n_2}\}_N]\nonumber\\
	&-\frac{1}{2}\mathop{\sum_{g_1,g_2}}_{g_1+g_2=g}\mathop{\sum_{m_1,m_2}}_{m_1+m_2=m}\mathop{\sum_{n_1,n_2}}_{n_1+n_2=n+2}\mathbf{ S}[\{ \mathcal{W}_{g_1,m_1,n_1}, \mathcal{W}_{g_2,m_2,n_2}\}_R]\nonumber\\
	&-\Delta_{NS} \mathcal{W}_{g-1,m+2,n}-\Delta_{R} \mathcal{W}_{g-1,m,n+2}
	\end{align} 
Here $\partial\mathcal{W}_{g,m,n}$ denotes the boundary of $\mathcal{W}_{g,m,n}$ and $\mathbf{S}$ denotes the operation of summing over inequivalent permutations of the external NS and also external R punctures.  $\{\mathcal{W}_{g_1,m_1,n_1}, \mathcal{W}_{g_2,m_2,n_2}\}_N$ denotes the set of Riemann surfaces with the choice of local coordinates and positions of picture changing-operators obtained by gluing the Riemann surfaces in $\mathcal{W}_{g_1,m_1,n_1}$ and $\mathcal{W}_{g_2,m_2,n_2}$ at one NS puncture from each via the {\it special plumbing fixture relation} 
\begin{equation}\label{specialplumbing}
zw=e^{i\theta} \qquad\qquad 0\leq\theta\leq2\pi
\end{equation}
where $z$ and $w$ denote the local coordinates around the punctures that are being glued. $\{\mathcal{W}_{g_1,m_1,n_1}, \mathcal{W}_{g_2,m_2,n_2}\}_R$ denotes the set of Riemann surfaces with the choice of local coordinates and distribution of PCO's obtained by gluing the Riemann surfaces in $ \mathcal{W}_{g_1,m_1,n_1}$ and $ \mathcal{W}_{g_2,m_2,n_2}$ at one R puncture from each of the surfaces via the special plumbing fixture relation. \par

Although definition of $\{\cdot,\cdot\}_N$ and $\{\cdot,\cdot\}_R$ looks the same, there is one additional subtlety in the definition of $\{\cdot,\cdot\}_R$. The total number of PCO's on the two Riemann surfaces corresponding to a point in $ \mathcal{W}_{g_1,m_1,n_1}$ and a point in $ \mathcal{W}_{g_2,m_2,n_2}$ is $$2(g_1+g_2)-4+(m_1+m_2)+\frac{n_1+n_2}{2}=2g-2+m+\frac{n}{2}-1$$ which is one less than the required number of PCO's on a Riemann surface with genus $g$, $m$ NS and $n$ R punctures. Therefore we need to prescribe the location of the additional PCO to define $\{\cdot;\cdot\}$.  A consistent prescription is to insert a factor of $$\chi_0=\oint \frac{dz}{z} \chi(z)$$ around one of the two punctures which are being glued, where $\chi(z)$ is the picture-changing operator.  Which of the two puncture that we choose is irrelevant since $$\oint \frac{dz}{z} \chi(z)=\oint \frac{dw}{w} \chi(z)$$ $\Delta_{NS}$ denotes the operation of taking a pair of NS punctures on a single Riemann surface corresponding to a point $ \mathcal{W}_{g-1,m+2,n}$ and gluing them via special plumbing fixture relation. Similarly, $\Delta_R$ denotes the operation of taking a pair of R punctures on a Riemann surface corresponding to a point $\mathcal{W}_{g-1,m+2,n}$ and gluing them via special plumbing fixture relation and in addition to this gluing we must also insert a factor of $\chi_0$ around one of the punctures.   

\subsection{Quantum BV master action}

In this subsection, we discuss the construction of the quantum BV master action for closed superstring field theory. Let us start by introducing the basic ingredients required for constructing the action. \par~\par

\noindent{\bf\underline{Important subspaces of $\mathcal{H}$} :} Consider the following subspaces of the full Hilbert space $\mathcal{H}$ of world-sheet SCFT:
\begin{itemize}

 \item $\mathcal{H}_T$ : The subspace of the Hilbert space $\mathcal{H}$ of world-sheet SCFT that is annihilated by $b_0^-$ and $L_0^-$. 
 
 \item $\mathcal{H}_{NS}$ :  The NS sector subspace of the heterotic string theory Hilbert space $\mathcal{H}_T$.
 
 \item $\mathcal{H}_R$ : The R sector subspace of the heterotic string theory Hilbert space $\mathcal{H}_T$.

 \item $\widehat{\mathcal{H}}_T$  : The subspace of $\mathcal{H}_T$ containing only NS sector states having picture number $-1$  and R sector states having picture number $-\frac{1}{2}$.
  
 \item $\widetilde{\mathcal{H}}_T$ : The subspace of  $\mathcal{H}_T$ containing only NS sector states  having picture number $-1$  and R sector states having picture number $-\frac{3}{2}$.
  
 \item $\widehat{\mathcal{H}}_+$ and $\widetilde{\mathcal{H}}_+$  : The subspaces that contain states in $\widehat{\mathcal{H}}_T$ and  $\widetilde{\mathcal{H}}_T$  of ghost numbers $\geq 3$.
 
 \item  $\widehat{\mathcal{H}}_-$ and $\widetilde{\mathcal{H}}_-$ : The subspaces that contain states in $\widehat{\mathcal{H}}_T$ and $\widetilde{\mathcal{H}}_T$ of ghost numbers $\leq 2$. 
 
 \end{itemize}

\noindent{\bf \underline{Space-time fields and antifields} :}  Let us expand the dynamical string field $|\Psi\rangle$  and the auxiliary string field  $|\widetilde\Psi\rangle$ as follows
\begin{align}\label{supersrtingfielde}
|\widetilde\Psi\rangle&=\sum_r|\widetilde\phi^r_-\rangle\widetilde\psi^r+\sum_r(-1)^{g_r^*+1}|{\widetilde\phi}^r_+\rangle\psi_r^*\nonumber\\
|\Psi\rangle-\frac{1}{2}\mathcal{G}|\widetilde\Psi\rangle&=\sum_r|\hat\phi^r_-\rangle\psi^r+\sum_r(-1)^{g_r^*+1}|\hat\phi^r_+\rangle\tilde\psi_r^*
\end{align} 
where  $g_r^*,g_r,\tilde g^*_r$ and $\tilde g_r$ are the Grassmann parities of $\psi^*_r, \psi_r,\tilde\psi^*_r$ and $\tilde \psi_r$ respectively.  The states $|\hat\phi^r_-\rangle, |\tilde\phi^r_-\rangle, |\hat\phi^r_+\rangle $ and $ |\tilde\phi^r_+\rangle $ are the basis states of  $\widehat{\mathcal{H}}_-,\widetilde{\mathcal{H}}_-, \widehat{\mathcal{H}}_+$ and $\widetilde{\mathcal{H}}_+$ satisfying the following orthonormality conditions
\begin{align}\label{orthonorm1}
\langle \hat\phi^-_r|c^-_0|\tilde\phi ^s_+\rangle&=\delta^s_r =\langle\tilde\phi^s_+|c^-_0|\hat\phi_r^- \rangle \nonumber\\
\langle \tilde\phi^-_r|c^-_0|\hat\phi^s_+ \rangle&=\delta^s_r =\langle\hat\phi^s_+|c^-_0|\tilde\phi _r^-\rangle
\end{align}
where $\langle\cdot|\cdot|\cdot\rangle$ denotes the BPZ inner product satisfying the following normalization
\begin{equation}
\langle p| c_{-1}\bar c_{-1}c_0\bar c_0c_1\bar c_1 e^{-2\phi(z)}|p'\rangle=(2\pi)^{-d}\delta^d(p+p')
\end{equation}
The target space objects $\{\psi_r,\tilde \psi_r\}$ are considered as the space-time fields and $\{\psi_r^*,\tilde \psi_r^*\}$ are considered as antifields.
 \par~\par

\noindent{\bf{\underline{Operator $\mathcal{G}$}} : }  Let us consider the operator $\mathcal{G}$  with the following action on the NS states and the R states:
\begin{equation} \label{hetg}
\mathcal{G}|s\rangle=
\begin{cases}
|s \rangle &\text{if}~|s\rangle\in \mathcal{H}_{NS}
\\
\chi_0 |s\rangle &\text{if}~|s\rangle\in \mathcal{H}_{R}
\end{cases}
\end{equation}
The operator $\chi_0$ is given by integrating the PCO over a cycle enclosing the puncture where we inserted the state $|s\rangle$ on the Riemann surfaces:
\begin{equation}\label{pcocycle}
\chi_0=\oint \frac{dz}{z} \chi(z)
\end{equation}
where $\chi(z)$ is the PCO defined as (\ref{hpart}). The operator $\mathcal{G}$ commutes with $Q_B$ and $b^{\pm}_0$:
\begin{equation}\label{chi0}
[Q_B,\mathcal{G}]=0\qquad \qquad [b^{\pm}_0,\mathcal{G}]=0
\end{equation}

\par~\par

 \noindent{\bf{\underline{Anti-linear bracket $\{\Psi^n\}_g$}} : } Consider the anti-linear bracket $\{\Psi,\cdots,\Psi\}_g$, with $n$ number of $\Psi's$ inside the bracket denoted as  $\{\Psi^n\}_g$. The  anti-linear bracket $\left\{A_1,\cdots,A_{m},B_1,\cdots,B_{n}\right\}_{g}$ associated with $m$ number of  off-shell NS states denoted by  $A_1,\cdots,A_{m}$  and $n$ number of off-shell R states denoted by $B_{1},\cdots,B_{n}$ is obtained by integrating the genus $g$ off-shell superstring measure associated with the off-shell states $A_1,\cdots,A_{m}, B_1,\cdots, B_n$ over the elementary string diagrams in the string vertex $\mathcal{W}_{g,m,n}$
 \begin{equation}\label{supstrinfvertex}
\left\{A_1,\cdots,A_{m},B_1,\cdots,B_{n}\right\}_{g}=\int_{\mathcal{MW}_{g,m,n}}\Omega^{g,m,n}_{D}\left(A_1,\cdots,A_{m},B_1,\cdots,B_{n}\right)
\end{equation}
where $D=6g-6+2m+2n$ and $\mathcal{MW}_{g,m,n}$ denotes the region inside the moduli space $\mathcal{M}_{g,m+n}$ covered by the string diagrams in the string vertex $\mathcal{W}_{g,m,n}$. The off-shell superstring measure $$\Omega^{g,m,n}_{D}\left(A_1,\cdots,A_{m},B_1,\cdots,B_{n}\right)$$ is built in two steps. First construct $\mathcal{B}_p$, the $p$-form component of the formal sum of operator valued differential forms,  given by

\begin{equation}
\mathcal{B}=\mathbf{K}\wedge \mathbf{B} \label{supstringmeasure1}
\end{equation}
where $\mathbf{K}$ is as follows
\begin{equation} 
\mathbf{K}=\sum_i G\left(z^i_1,\cdots,z^i_N\right)\left(\chi(z^i_1)-\partial \xi(z^i_1)dz^i_1\right)\wedge\cdots\wedge\left(\chi(z^i_{N})-\partial \xi(z^i_{N})dz^i_{N}\right) \label{krforms}
\end{equation}
with $N=2g-2+m+\frac{n}{2}$. The sum over $i$ denotes the sum over different PCO configurations such that $\mathbf{K}$ becomes the mapping class group invariant PCO distribution assigned to the elementary string diagrams belong to the string vertex $\mathcal{W}_{g,m,n}$. The function  $G\left(z^i_1,\cdots,z^i_N\right)$ is a weight factor for each PCO configuration in the total distribution satisfying 
\begin{equation} 
\sum_i G\left(z^i_1,\cdots,z^i_N\right)=1 \label{krforms1}
\end{equation}
The operator valued differential form $\mathbf{B}$ is built using the beltrami differentials $\left(\mu_1,\cdots, \mu_D\right)$ on a Riemann surface. These beltrami differentials generate a deformation in the complex structure on a Riemann surface that correspond to the coordinates $\left(t_1,\cdots,t_D\right)$ on the moduli space. Then the form $\mathbf{B}$  is given by
\begin{equation}
\mathbf{B}=\prod_{k=1}^D \left\{ 1+b(\mu_k)\delta t_k \right\}
\end{equation}
where $b(\mu_k)$ is obtained by integrating the contraction of the beltarmi differential $\mu_k$ with the reparameterization field $b$  over the world-sheet:
\begin{equation}
b(\mu_k)=\int d^2z\left( b_{zz}\mu^z_{k\bar z}+b_{\bar z\bar z}\mu^{\bar z}_{k z}\right)
\end{equation}
The second step is to construct the surface state $|\Sigma\rangle$ associated with the string diagram $\mathcal{R}$, which describes the state that is created on the boundaries of $D_i$ by performing a functional integral over the fields of SCFT on $\mathcal{R}-\sum_i D_i$. Then we can obtain the off-shell superstring measure as the following expectation value
\begin{equation}
\Omega^{g,m,n}_{D}\left(A_1,\cdots,A_{m},B_1,\cdots,B_{n}\right)=\left(2\pi\mathbf{i}\right)^{-3g+3-m-n}\langle \Sigma|\mathcal{B}_D|A_1\rangle\otimes\cdots\otimes|A_{m}\rangle|B_1\rangle\otimes\cdots\otimes|B_{n}\rangle
\end{equation}
The inner product between $\langle\Sigma|$ and the state $$|A_1\rangle\otimes\cdots\otimes|A_{m}\rangle|B_1\rangle\otimes\cdots\otimes|B_{n}\rangle\in\mathcal{H}_{NS}^{\otimes m}\otimes\mathcal{H}_{R}^{\otimes n}$$  describes the $m+n$-point correlation function on $\mathcal{R}$ with  vertex operators associated with off-shell states inserted at the punctures using the local coordinate choices around that puncture assigned to the elementary string diagrams in the string vertex $\mathcal{W}_{g,m,n}$. \par~\par

\noindent{\bf{\underline{Quantum BV action}} :} The quantum master action for the superstring field theory \cite{Sen:2015uaa}  is given by  
\begin{equation}\label{supstringfieldqmaction}
S=g_s^{-2}\left[-\frac{1}{2}\langle\widetilde\Psi|c_0^-Q_B\mathcal{G}|\widetilde\Psi \rangle+\langle\widetilde\Psi|c_0^-Q_B|\Psi \rangle+\sum_{g=0}^{\infty}(\hbar g_s^2)^g\sum_{n=1}^{\infty}\frac{g_s^n}{n!}\{\Psi^n\}_g\right]
\end{equation}
where $g_s$ denotes the closed string coupling. \par ~\par

\noindent{\bf{\underline{Gauge fixing}} : } Given the master action, we compute the quantum amplitudes by carrying out the usual path integral over a Lagrangian submanifold of the full space spanned by  $\psi^r$ and $\psi^*_r$. The Lagrangian submanifold can be specified by the Siegel gauge condition

\begin{equation}\label{siegelgsup}
b_0^+|\Psi\rangle=0,~ b_0^+|\widetilde\Psi\rangle=0 \quad\Rightarrow\quad b_0^+\left(|\Psi\rangle-\frac{1}{2}\mathcal{G}|\widetilde\Psi\rangle\right)=0
\end{equation}
In this gauge, the propagator in $|\widetilde\Psi\rangle$, $|\Psi\rangle$ space takes the following form 

\begin{equation}\label{supprop}
-g_s^2\frac{b_0^+b_0^-\delta_{L_0,\bar L_0}}{L_0+\bar L_0}\left(\begin{array}{cc}0 & 1 \\1 & \mathcal{G}\end{array}\right)
\end{equation}

We discussed in subsection \ref{dsupstringfield} that the dynamical and auxiliary string are is built using NS  states in $-1$ picture and R  states in  either $-\frac{1}{2}$ or $-\frac{3}{2}$ picture. We already discussed that except for the -1 picture, there are infinite number of NS states at any fixed mass$^2$ level. Hence, allowing only the $-1$ picture avoids the infinity due to infinite number of particles propagating inside  loops. Unfortunately, there are infinite number of R states even in the $-\frac{1}{2}$ and $-\frac{3}{2}$ pictures  at any mass$^2$ level. In the $-\frac{1}{2}$ picture sector, these infinite number of states are created by the action of $\gamma_0$ oscillator and in the conjugate $-\frac{3}{2}$ picture sector, they are created by the action of $\beta_0$ oscillator. Fortunately, the R sector propagator 
\begin{equation}
\langle \phi_s^c|c_0^-b_0^+(L_0^+)\chi_0|\phi_r^c \rangle
\end{equation}
allows  only finite number of states out of these infinite number of states to propagate.  In order to see this, consider the infinite number of states created by the action of $\beta_0$ oscillators on a $-\frac{3}{2}$ picture state. Since $\beta_0$ has ghost number $-1$, these states will have arbitrarily low ghost numbers. By acting these states with $\chi_0$, we can obtain states of picture number $-\frac{1}{2}$ and arbitrarily low ghost numbers. But such states do not exists in the $-\frac{1}{2}$  picture sector. In the $-\frac{1}{2}$  picture sector at a fixed mass$^2$ level, we can only have states with arbitrarily large ghost numbers created by $\gamma_0$ oscillators. This implies that the $\chi_0$ factor must annihilate all except a finite number of states created by the $\beta_0$ oscillators \cite{deLacroix:2017lif}. \par

\section{Construction of String Vertices using Hyperbolic  Surfaces}\label{cons}

 In this section, we shall discuss the construction of string vertices in closed superstring field theory using Riemann surfaces endowed with metric having constant curvature $-1$, commonly known as hyperbolic Riemann surfaces.

\subsection{Fuchsian representation for hyperbolic Riemann surfaces }

Intuitively,  a Riemann surface $\mathcal{R}$ can be understood as a collection of domains in a complex plane which are glued together by biholomorphic mappings. This collection of domains and the biholomorphic mappings is called the complex structure of the Riemann surface. Two surfaces constructed using two different collections of domains and biholomorphic mappings represent the same Riemann surface if the domains in one collection can be obtained by a biholomorphic mapping  of the domains in the other collection.  For instance, it is known that an arbitrary simply connected Riemann surface is biholomorphically equivalent to one of the  three following Riemann surface:  complex plane, Riemann sphere and  upper half plane. However, there are infinitely many different multiply connected Riemann surfaces. Interestingly, all of them can be constructed from simply connected Riemann surfaces using the notion of the universal covering of a Riemann surfaces. \par

Let $\mathcal{R}$ be a multiply connected Riemann surface and $\widetilde{\mathcal{R}}$ be a simply connected Riemann surface. Then it is always possible to find a many to one map 
\begin{equation}
\pi : \widetilde{\mathcal{R}}\to \mathcal{R}
\end{equation}
 known as the universal covering map, that is biholomorphic  if we restrict the map $\pi$ to a connected region $\widetilde u$ of the inverse image $\pi^{-1}(u)$ of the neighbourhood $u$ of a point $p$ in $\mathcal{R}$. This means that two different points  on $\widetilde{\mathcal{R}}$ connected via a biholomorphic mapping $\gamma$ might be representing the same point on $\mathcal{R}$, i.e.
 \begin{equation}
 \pi \circ \gamma=\pi
 \end{equation} 
Such a transformation $\gamma$ is known as a universal covering transformation. All such covering transformations together form the universal covering group $\Gamma$ of $\mathcal{R}$. Then we can obtain the Riemann surface $\mathcal{R}$ by considering the quotient of the simply connected Riemann surface $\widetilde{\mathcal{R}}$ under the action of group $\Gamma$:
\begin{equation}
\mathcal{R}=\widetilde{\mathcal{R}}/\Gamma
\end{equation}
For example, we can obtain the puncture complex plane $\mathbb{C}-\{0\}$ from the complex plane $\mathbb{C}$ by identifying points on $\mathbb{C}$ under the action of group generated by the transformation $$z\to z+2\pi \mathbf{i} \qquad\qquad n \in \mathbb{Z}$$ The associated covering map $$\pi : \mathbb{C}\to \mathbb{C}-\{0\}$$ is given by $$\pi(z)=e^{z}$$

Now we shall consider the multiply connected Riemann surface whose universal covering space is the upper half-plane $\mathbb{H}$. Let us introduce the Poincar\'e metric on $\mathbb{H}$, given by
\begin{equation}\label{hdisk}
ds^2_{\mathbb{H}}=\frac{|dz|^2}{\left(\text{Im} z\right)^2}
\end{equation}
The Gaussian curvature, which is the half of the Ricci curvature, of this metric is $-1$ everywhere on $\mathbb{H}$.  The geodesics on $\mathbb{H}$ with respect to the Poincar\'e metric are either an arc that belongs to the half circle on $\mathbb{H}$ with centre on the real axis or a straight line parallel to the imaginary axis.  The transformation on $\mathbb{H}$ 
\begin{equation}
\gamma(z)=\frac{az+b}{cz+d} \qquad\qquad a,b,c,d\in \mathbb{R}\qquad ad-bc=1
\end{equation}
 preserve the Poincar\'e metric. All such transformations together form the isometry  group  $PSL(2,\mathbb{R})$.  \par
 
 Let us consider a discrete subgroup $\Gamma$ of $PSL(2,\mathbb{R})$. It is straight forward to see that the quotient of the upper half-plane endowed with the Poincar\'e metric with respect to the group $\Gamma$  is a hyperbolic Riemann surface $\mathcal{R}$ with Gaussian curvature $-1$  all over the surface:
 \begin{equation}
 \mathcal{R}=\mathbb{H}/\Gamma
 \end{equation}
  The quotient $\mathbb{H}/\Gamma$ is known as the {\it Fuchsian representation} of $\mathcal{R}$ and $\Gamma$ is called as the Fuchsian group associated with $\mathcal{R}$. In fact, the {\it Fuchsian uniformaization theorem} due to Poincar\'e guarantee  that for any hyperbolic Riemann surface the universal covering space is the Poincar\'e upper half-plane $\mathbb{H}$ and the universal covering group is a discrete subgroup of $PSL(2,\mathbb{R})$. \par 
  
  For example, the hyperbolic thrice punctured sphere is obtained by considering the quotient of $\mathbb{H}$ with respect to the group $\Gamma(2)$ generated by the transformations $$z\to \frac{z}{2z+1}\qquad \qquad z\to z+2$$

\subsection{Elementary string diagrams and Hyperbolic surfaces }

Since the goal is to construct the elementary string diagrams building the string vertices, we need to check whether  the representative of a Riemann surface which we consider have all the desired features.\par~\par

\noindent{\bf{\underline{Symmetric local coordinates} :}} The Riemann surface representing an elementary string diagram must have local coordinates around the punctures that is independent of the labelling of the punctures. Interestingly, the natural local coordinates around the punctures on a hyperbolic Riemann surface has this property. To see this, consider a puncture $p$ on a hyperbolic Riemann surface. There is a distinguished local conformal coordinate with $w(p)=0$ and the metric locally given by \cite{SAWolpert1}
\begin{equation}\label{cusp}
ds^2=\left(\frac{|dw|}{|w|\text{ln}|w|}\right)^2
\end{equation}
The canonical coordinate $w$ is unique up to a phase factor. The circle $|w| = r$ is the closed horocycle about $p$ with hyperbolic length $$l(r) = -\frac{2\pi}{\text{ln}r}$$ The  hyperbolic metric for the punctured unit disc $D = \{0 < |w| < 1\}$ is also given by the formula (\ref{cusp}).  It is a well known fact  that the unit area neighbourhood of the puncture $|w| \leq  r$ with $l(r) \leq 1 $ isometrically embeds to a neighbourhood of each puncture on a hyperbolic Riemann surface \cite{L}. As a result, the local coordinates around the punctures on a Riemann surface induced from the hyperbolic metric is independent of the labelling of the puncture.\par~\par

\noindent{\bf{\underline{Gluing compatible local coordinates} :}} The geometric identity (\ref{bvmastercond1}) demands that the local coordinates on the representative Riemann surfaces which we declare as elementary string diagrams must have the following continuity property, known as gluing compatibility, across the boundary of the string vertices inside the moduli space. A consistent set of string vertices together with the string diagrams obtained by gluing elementary string diagrams that belong to the string vertices cover the moduli space once. It is possible only if the data encoded on the representative  Riemann surface which are identified as elementary string diagrams must match across the boundaries of the string vertices inside the moduli space. This means that the representative Riemann surfaces that belongs to the boundary of the string vertices must have local coordinates around the punctures that agrees up to a phase factor with that on the Riemann surfaces obtained by gluing the representative Riemann surfaces using the special plumbing fixture relation (\ref{specialplumbing}). \par

From the above discussion, it is clear that hyperbolic Riemann surfaces can be used to build string vertices only if the natural local coordinates with metric (\ref{hdisk}) around the punctures is gluing compatible.  For this, we must  analyze the plumbing fixture of hyperbolic Riemann surfaces in detail. This has been already done in \cite{SAWolpert2}. Unfortunately, the natural local coordinates around the punctures  induced from the hyperbolic metric is not gluing compatible. However, the good news is that if the length of the core geodesic on the glued surface is very small compared to the geodesic that passes through the core geodesic perpendicular, then the difference between the local coordinates on the glued surface and the canonical local coordinates  on a hyperbolic Riemann surface vanishes to first order in the length of the core geodesic on the glued surface. Using this fact a continuous choice of modified local coordinates on the string diagrams inside the string vertices has been proposed in \cite{Moosavian:2017qsp}.  \par

Following \cite{Moosavian:2017qsp}, we define the string vertices in closed superstring field theory using hyperbolic Riemann surfaces as follows.\par~\par

\noindent{\bf{\underline{String vertex $\mathcal{W}_{g,m,n}$} :}} Consider hyperbolic Riemann surfaces with $g$ number of handles and $m+n$ number of punctures having no simple closed geodesics with length less than  $c_*$. Then, the string vertex $\mathcal{W}_{g,m,n}$ is defined as the set of all such inequivalent Riemann surfaces having a {\it consistent choice of distribution}  of $2g-2+m+\frac{n}{2}$ number of PCO's and  following  local coordinates  around the punctures.  

\begin{itemize}
\item  The local coordinate around the $j^{th}$ puncture on  Riemann surfaces having no simple closed geodesics with length less than $(1+\delta)c_*$ is  $e^{\frac{\pi^2}{c_*}}w_j$, where $w_J$ is natural local coordinate induced from the hyperbolic metric, and $c_*,\delta$ are arbitrary infinitesimal real parameters.
\item  The local coordinate around the $j^{th}$ puncture on Riemann surfaces with  $k$ number of degenerating disjoint simple closed geodesics having length in between $c_*$ and $(1+\delta)c_*$ is  $e^{\frac{\pi^2}{c_*}}\hat w_j$, where $\hat w_j$ is given by
\begin{equation}
\left(\frac{|d\hat w_j|}{|\hat w_j|\text{ln}|\hat w_j|}\right)^2=\left(\frac{|dw_j|}{|w_j|\text{ln}|w_j|}\right)^2\left\{1+\frac{c_*^2}{3~\text{ln}|w_j|}\sum_{i=1}^kf(l_i)Y_{ij}\right\}
\end{equation}
 
\end{itemize}
$l_i$ denotes the length of the $i^{th}$ degenerating simple closed geodesic and the function $f(l_i)$ is an arbitrary smooth real function of the geodesic length $l_i$ defined in the interval $\left(c_*,c_*+\delta c_*\right)$, such that $f(c_*)=1$ and $f(c_*+\delta c_*)=0$. The coefficient $Y_{ij}$ is  the leading order term in the following sum around the $j^{th}$ puncture  
\begin{equation}\label{mlocal}
 \text{ln}|w_j|\left(E_{i,1}+E_{i,2}\right)
\end{equation}
where,   $E_{i,1}, E_{i,2}$ denote the Eisenstein series associated with the cusps that are being glued via plumbing fixture to get the collar whose core geodesic is the $i^{th}$ degenerating simple closed geodesic.   The definition and  the expansion of Eisenstein series around a cusp is discussed in appendix (\ref{Eisen}). By using the results discussed there, we obtain $Y_{ij}$ as follows

\begin{align}\label{yij}
Y_{ij}&=\sum_{q=1}^2\sum_{c_i^q,d_i^q}\pi^{2}\frac{\epsilon(j,q)}{|c_i^q|^4}\nonumber\\ c_i^q>0 \qquad &d_i^q~\text{mod}~c_i^q \qquad\left(\begin{array}{cc}* & * \\c_i^q & d_i^q\end{array}\right)\in \quad (\sigma_i^q)^{-1}\Gamma_{i}^{q}\sigma_j
\end{align} 
Here, $\Gamma_i^q$ denotes the Fuchsian group for the component Riemann surface with the cusp  denoted by the index $q$ that is being glued via plumbing fixture to obtain the $i^{th}$ collar.  The transformation $\sigma_j^{-1}$ maps the cusp corresponding to the $j^{th}$ cusp to $\infty$ and $(\sigma_j^q)^{-1}$ maps the cusp denoted by the index $q$ that is being glued via plumbing fixture to obtain the $i^{th}$ collar to $\infty$. The factor $\epsilon(j,q)$ is one if both the $j^{th}$ cusp and he cusp denoted by the index $q$ that is being glued via plumbing fixture to obtain the $i^{th}$ collar belong to the same component surface other wise $\epsilon(j,q)$ is zero.

\subsection{One-loop tadpole string vertex}

   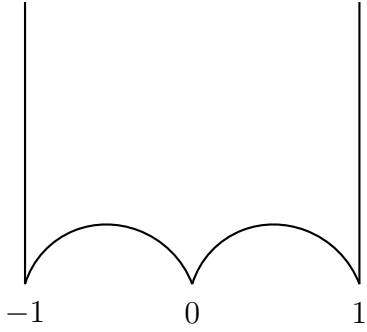
\begin{figure}
\begin{center}
\usetikzlibrary{backgrounds}
\begin{tikzpicture}[scale=.55]
\draw[black,  thick] (2.3,.2) to[curve through={(3.5,1.5)..(5,1.5)}] (6.3,.2);
\draw[black,  thick] (6.3,.2) to[curve through={(7.5,1.5)..(9,1.5)}] (10.3,.2);
\draw[black,  thick] (2.3,.2)--(2.3,7);
\draw[black,  thick] (10.3,.2)--(10.3,7);
\draw  node at (10.3,-.5) {$1$};
\draw  node at (2.3,-.5) {$-1$};
\draw  node at (6.3,-.5) {$0$};
\end{tikzpicture}
\end{center}

\caption{Fundamental domain for a thrice-punctured sphere. }
\label{Fundamental domain for thrice-punctured sphere}
\end{figure}

Let us elucidate the definition of string vertex described above by explicitly constructing the one-loop tadpole string vertex. An elementary string diagram in the one-loop tadpole vertex has one puncture and one handle.  The moduli space of once-punctured tori has only one boundary, and it corresponds to once-punctured tori obtained via the plumbing fixture of thrice-punctured sphere.\par

 The Fuchsian group for the thrice-punctured sphere is the modular group $\Gamma(2)$ generated by the matrices $ \left(\begin{array}{cc}1 & 0 \\2 & 1\end{array}\right)$ and $\left(\begin{array}{cc}1 & 2 \\0 & 1\end{array}\right) $. A fundamental domain of $\Gamma(2)$ can be represented by a hyperbolic rectangle on the upper half-plane with corners at $0,1,-1$ and $\text{i}\infty$, as shown in figure (\ref{Fundamental domain for thrice-punctured sphere}). Let us denote the transformations that map $0,1$ to $\text{i}\infty$ as $\sigma_0,\sigma_1$ respectively and are given by $$\sigma_0=\left(\begin{array}{cc}0 & 1 \\-1 & 0\end{array}\right)\qquad \qquad \sigma_1=\left(\begin{array}{cc}1 & 0 \\1 & -1\end{array}\right)$$ 
Assume that the degenerate once-punctured torus is obtained by the gluing of cusps at $0$ and $1$.  In this situation, the series (\ref{yij}) takes the following form 
\begin{align}\label{Y}
Y=\pi^2\sum_{i=1}^2\sum_{m_i}\frac{\chi_i(m_i)}{|2m_i+1|^4}
\end{align} 
where, $\chi_1(m_1)$ is the number of representatives  $\left(\begin{array}{cc}2p_1 & 2q_1+1 \\2m_1+1 & 2n_1\end{array}\right)$ of distinct double cosets in $\Gamma_0\backslash \sigma_0\Gamma(2)/\Gamma_0$ with a fixed $m_1\geq 0$. The group $\Gamma_0$ is the set of all matrices of the form  $\left(\begin{array}{cc}1 & 2b \\0 & 1\end{array}\right)$ with $b\in \mathbb{Z}$. Similarly, $\chi_2(m_2)$ is the number of representatives  $\left(\begin{array}{cc}2p_2+1 & 2q_2 \\2m_2+1 & 2n_2+1\end{array}\right)$ of distinct double cosets in $\Gamma_0\backslash \sigma_1\Gamma(2)/\Gamma_0$ with a fixed $m_2\geq 0$.\par~\par

With this background, we can define the one-loop tadpole string vertex as follows.\par~\par
\noindent{\bf{\underline{One-loop tadpole string vertex $\mathcal{W}_{1,1,0}$} :}} Consider once-punctured hyperbolic  tori  having no simple closed geodesics with length less than  $c_*$. Then, the string vertex $\mathcal{W}_{1,1,0}$ is defined as the set of all such inequivalent Riemann surfaces having a {\it consistent distribution}  of a PCO and  the following  local coordinate  around the puncture.
\begin{itemize}
\item  The local coordinate around the puncture on  a once-punctured torus having no simple closed geodesics with length less than $(1+\delta)c_*$ is  $e^{\frac{\pi^2}{c_*}}w$, where $w$ is natural local coordinate induced from the hyperbolic metric, and $c_*,\delta$ are arbitrary infinitesimal real parameters.
\item  The local coordinate around the puncture on a once punctured torus with a  simple closed geodesics having length in between $c_*$ and $(1+\delta)c_*$ is  $e^{\frac{\pi^2}{c_*}}\hat w$, where $\hat w$ is given by
\begin{equation}
\left(\frac{|d\hat w|}{|\hat w|\text{ln}|\hat w|}\right)^2=\left(\frac{|dw|}{|w|\text{ln}|w|}\right)^2\left\{1+\frac{c_*^2}{3~\text{ln}|w|}Yf(l)\right\}
\end{equation}
where, $Y$ is given by (\ref{Y}) and  $f(l)$ is an arbitrary smooth real function of the length $l$ of the geodesic on the degenerating collar defined in the interval $\left(c_*,c_*+\delta c_*\right)$, such that $f(c_*)=1$ and $f(c_*+\delta c_*)=0$. 
\end{itemize}

\section{Consistent Choice of PCO Distribution }

The definition of string vertices is complete only if we are able to specify a consistent choice of PCO distribution. The PCO distribution must have the following properties:
\begin{itemize}
\item symmetric with respect to the punctures;
\item invariant under mapping class group transformations;
\item compatibility with the special plumbing fixture.
\end{itemize}
 In this section, we describe a systematic procedure for constructing a choice of PCO distribution having all the  above mentioned properties.

\subsection{PCO distribution on once-punctured tori}

Consider a once-punctured torus with a $-1$ picture NS state  inserted at the puncture. Since R punctures come only in pairs, we are only allowed to insert  an NS state at the puncture.  The vanishing of picture number requires distributing a PCO on the surface.  One might attempt to insert a PCO at a point on the surface  by specifying a point in the fundamental region for the once punctured torus in the Poincar\'e upper half-plane.  As a matter of fact, such a prescription does not go well due with the large diffeomorphisms acting on the surface.  There  are infinitely many different points in the upper half-plane representing the same point on a once-punctured torus.  The nice feature of this set of infinite points is that, it is possible to reach any of these infinite number of points from one point in the set by acting with an element in the mapping class group, the group of all large diffeomorphisms, of the once-punctured torus. Therefore, by placing a PCO at all these points  on the upper half-plane and by considering the average  we can obtain a distribution of PCO that is unaffected by the action of large diffeomorphisms. \par
   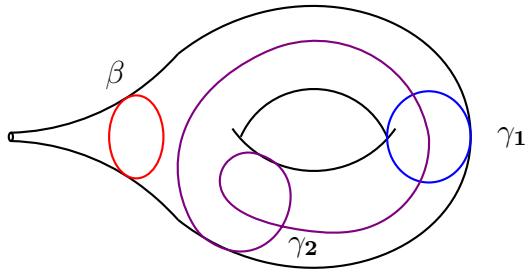
\begin{figure}
\begin{center}
\usetikzlibrary{backgrounds}
\begin{tikzpicture}[scale=.55]

\draw[black,   thick] (1,2) to[curve through={(3,3)..(8,0) .. (3,-3)}] (1,-2);
\draw[black,   thick] (1,2) to[curve through={(.5,1.5)..(0,1.1)}] (-3,0.1);
\draw[black,   thick] (1,-2) to[curve through={(.5,-1.5)..(0,-1.1)}] (-3,-.1);
\draw[black,  thick] (2.5,0) to[curve through={(3.5,1)..(5,1)}] (6,0);
\draw[black,  thick] (2.3,.2) to[curve through={(3.5,-.7)..(5,-.7)}] (6.2,.2);
\draw[black, thick](-3,0) ellipse (.05 and .1);
\draw[blue,   thick](7,0) ellipse (1 and 1.1);
\draw[red,   thick](0,0) ellipse (.65 and 1);
\draw[violet,  thick, ] (7,0) to [curve through={(3.5,2.2)..(3,2)..(1.1,0)..(1.3,-2)..(1.65,-2.45)..(2.85,-.4)..(2,-1.1)..(2.7,-1.9)..(4,-2.25)..(5.5,-2.2)}](7,0);
\draw  node at (9,0) {$\mathbf{\gamma_1}$};
\draw  node at (-.5,1.5) {$\mathbf{\beta}$};
\draw  node at (4,-2.65) {$\mathbf{\gamma_2}$};
\end{tikzpicture}
\end{center}

\caption{Different classes of simple closed curves on a once-punctured hyperbolic torus. The curves $\gamma_1,\gamma_2$ are simple closed geodesics and the  curve $\beta$ is not a geodesic.  }
\label{geodesics on tori}
\end{figure}
Unfortunately, such a naive PCO distribution is not well defined, since an arbitrary sum over infinite number of points often fails to converge. However, by considering a weighted average instead of a simple average we can make it a sensible PCO distributions. For this, assume that there exist an identity of the following form:
\begin{equation}\label{identityMCG}
\sum_{h\in \text{MCG}(\mathcal{R}_{1,1})} G( h\cdot\gamma)=Q
\end{equation} 
where, $Q$ is some constant  and the sum is over the action of all elements in the mapping class group MCG$(\mathcal{R}_{1,1})$ on a simple closed geodesic $\gamma$ on the once puncture torus $\mathcal{R}_{1,1}$. The summand $G(\gamma)$ is a function of quantities associated with the simple closed geodesic $\gamma$. Note that such simple closed geodesic on a once punctures torus is not homotopic to the puncture.  Moreover, there is no simple closed geodesic on a once-punctured simple closed curves of the type $\beta$ shown in figure  (\ref{geodesics on tori}) are homotopic to the puncture. The simple closed geodesic $\gamma_2$ is the image of the action of a mapping class group element on the simple closed geodesic $\gamma_1$.  Then a PCO distribution of the following kind is manifestly invariant under the action of mapping class group and well defined:
\begin{equation}\label{MCGPCO}
\sum_{h\in \text{MCG}(\mathcal{R}_{1,1})} G( h\cdot\gamma) \chi(\gamma\cdot p)
\end{equation} 
 where, $\chi(P)$ denotes a PCO inserted at a point $P$ in the fundamental region of a once punctured torus in $\mathbb{H}$. \par
 
 Let us analyze the properties that the PCO distribution (\ref{MCGPCO}) must have in order to consider it as a consistent PCO distribution on  once punctured tori that make up the one-loop tadpole string vertex. First of all, if a simple closed geodesic on a once punctured tori have length $c_*$ then the PCO distribution on it must match with the PCO distribution induced on a once punctured torus constructed by the plumbing fixture of two puncture on a thrice punctured sphere, with the core geodesic on the plumbing collar having length $c_*$. This is the gluing compatibility requirement.  \par~\par
  \begin{figure}
\begin{center}
\usetikzlibrary{backgrounds}
\begin{tikzpicture}[scale=.5]
\draw[ thick] (-30,-6)--(-10,-6);
\draw node at (-9,-6) {$R$};
\draw[black,thick] (-22,-5) to[curve through={(-20,-4)}] (-18,-5);
\draw[black,thick] (-24,-4) to[curve through={(-22.5,-4.5)}] (-22,-5);
\draw[black,thick] (-27,-2) to[curve through={(-24.5,-3)}] (-24,-4);
\draw[black,thick] (-28,0)to[curve through={(-27.5,-.5)}] (-27,-2) ;
\draw[black,thick] (-28,0)to[curve through={(-20,4.25)}] (-12,0) ;
\draw[black,thick] (-13,-2)to[curve through={(-12.5,-.5)}] (-12,0) ;
\draw[red,thick] (-13,-2)to[curve through={(-15.5,-3)}] (-16,-4) ;
\draw[black,thick] (-16,-4) to[curve through={(-17.5,-4.5)}] (-18,-5);
\draw[red,thick] (-25,-2.5) to[curve through={(-20,-.1)}] (-15,-2.5);
\draw[red,thick] (-18.75,-4.4) to[curve through={(-14,-.82)}] (-12.7,-.81);
\draw[red,thick] (-17,-4.2) to[curve through={(-14.65,0)}] (-12.72,1);
\draw[red,thick] (-21.25,-4.4) to[curve through={(-24.5,-.82)}] (-27.3,-.81);
\draw[red,thick] (-23,-4.2) to[curve through={(-24.65,0)}] (-27.2,1);
\draw node [above] at (-20,4.25) {$\mathlarger{\mathlarger{L_3}}$};
\draw node [below,right] at (-15.5,-3) {$\mathlarger{\mathlarger{L_1}}$};
\draw node [below,left] at (-24.5,-3) {$\mathlarger{\mathlarger{L_1}}$};
\draw node  [below] at(-20,-4) {$\mathlarger{\mathlarger{L_2}}$};
\draw[blue,thick](-20,4.25)--(-20,-4);
\draw[red,thick](-23.6,-1.2) ellipse (.1 and .1);
\draw[red,thick](-16.025,-1.475) ellipse (.1 and .1);
\draw node [right ] at (-24,-.5) {\color{black}$\mathlarger{\mathlarger{p_1}}$};
\draw node  at (-16.5,-.6) {\color{black}$\mathlarger{\mathlarger{p_2}}$};
\end{tikzpicture}
\end{center}
\end{figure}
\noindent{\bf{\underline{Gluing of NS punctures} :}} Consider the situation where two NS punctures of a thrice NS punctured sphere are glued to obtain a once-punctured torus with one NS puncture. On a thrice puncture sphere three $-1$ picture NS states at the punctures, we must insert a  PCO.  A  consistently distributed PCO on a thrice punctured sphere must be symmetric with respect to the punctures. Such a distribution of a PCO can be obtained by considering {\it the average of a PCO placed at each of the centroids of the two hyperbolic triangles obtained by scissoring the thrice punctured hyperbolic sphere along the three disjoint geodesics that connect a pair of punctures on it}:
 \begin{equation}
 \frac{1}{2}\left\{\chi(p_1)+\chi(p_2)\right\}
 \end{equation}
 where, $p_1$ and $p_2$ denote the  centroids of the hyperbolic triangles that are glued together to obtain the thrice punctured sphere. This prescription provide a PCO distribution on a thrice punctured sphere that is not only symmetric with respect to the punctures, it is also invariant under the {\it involution symmetry} of the thrice-punctured hyperbolic sphere. The involution symmetry states that the exchange of the hyperbolic triangles that are glued together to obtain the thrice punctured sphere does not change the surface. \par
 
 Let us modify the PCO  distribution (\ref{MCGPCO}) to  the following distribution 
\begin{equation}\label{MCGPCO1}
\sum_{h\in \text{MCG}}  \frac{1}{2}G( h\cdot\gamma) \left\{\chi( p_1^{h\cdot \gamma})+\chi( p_2^{h\cdot \gamma})\right\}
\end{equation}
where, $p_1^{\gamma}$ and $p_2^{\gamma}$ denote the centroids of the pentagons that are glued together to construct the surface obtained by the scissoring of once-punctured torus along the simple closed geodesic $\gamma$. This PCO distribution is compatible with the gluing of two NS punctures, provided the function $G$ has the following properties:

\begin{align}\label{GPCO1}
 \lim_{l_{\gamma}\to c_*} \sum_{h\in \text{Dehn}(\gamma)} G( h\cdot\gamma)&=1+\mathcal{O}\left(c_*^3\right) \nonumber\\
 \lim_{l_{\gamma}\to c_*}G( g\cdot\gamma)&=\mathcal{O}\left(e^{-1/c_*}\right)\qquad\qquad \forall \quad g\in \text{MCG}(\mathcal{R}_{1,1})/\text{Dehn}(\gamma)
\end{align}
  where, $c_*$ is same as the infinitesimal real parameter introduced in the previous section for defining the string vertices, and  $\text{Dehn}(\gamma)$ denotes the subgroup of mapping class group that is generated by the  Dehn twist performed with respect to the simple closed geodesic $\gamma$. To check the gluing compatibility of the PCO distribution (\ref{MCGPCO1}), let us assign this PCO distribution on a once-punctured torus with simple closed geodesic length having infinitesimal length $c_*$. Using the limiting behaviour of the function $G$ (\ref{GPCO1}) we can check that it takes the following form
  \begin{equation}\label{MCGPCO2}
 \frac{1}{2}\left\{\chi( p_1^{ \gamma})+\chi( p_2^{ \gamma})\right\}+\mathcal{O}\left(c_*^3\right)
\end{equation}

It is clear that the proposed PCO distribution (\ref{MCGPCO1}) is gluing compatible, if the locations $p_1^{\gamma}, p_2^{\gamma}$ of the centroids of the hyperbolic pentagons matches with $p_1^{c_*}, p_2^{c_*}$ the locations of the centroids $p_1, p_2$ of the hyperbolic triangles on the surface with core geodesic having length $c_*$ on the plumbing collar constructed via the plumbing fixture. The locations  $p_1^{c_*}, p_2^{c_*}$ are also the centroids of  hyperbolic pentagons.   Note that, one of the corners of all these pentagons touches the real axis of $\mathbb{H}$.  We can obtain such a hyperbolic pentagon from a  hyperbolic hexagon by taking the length of one the sides to zero by shrinking the corresponding edge to a point in the real axis. It is a well known fact in hyperbolic geometry that a hyperbolic hexagon is uniquely determined by the lengths of the three non-adjacent sides. Moreover, the location of the centroid of a hyperbolic hexagon is also uniquely determined by the lengths of the three non-adjacent sides. Therefore, the location of the centroid of any hyperbolic pentagon on the upper half-plane can be specified as a function of the  lengths of the two sides that determines the pentagon.\par

  It is straight forward to see that the hyperbolic pentagon obtained by scissoring a once-punctured torus with simple geodesic $\gamma$ having length $c_*$ has the same length $c_*/2$ for both the edges that uniquely determine the pentagon. A careful analysis of the plumbing fixture of hyperbolic Riemann surfaces  reveals that the resulting surface  have an induced metric that deviates from the hyperbolic metric. Consider the Riemann surface obtained by the gluing of two punctures on a thrice-punctured sphere with plumbing collar having core geodesic with length $c_*$. The length of the core geodesic on the plumbing collar computed using the hyperbolic metric on the glued surface has a deviation from $c_*$ that is only of the order $c_*^3$ \cite{Wolpert1}.  As a result, that both set of locations $p_1^{\gamma}, p_2^{\gamma}$ and $p_1^{c_*}, p_2^{c_*}$ matches up to $\mathcal{O}(c_*^2)$. Hence, the PCO distribution (\ref{MCGPCO1}) is compatible with the gluing of NS punctures up to $\mathcal{O}(c_*^2)$, provided there exist an identity of the type (\ref{identityMCG}) with the properties (\ref{GPCO1}).\par

 Therefore, our next goal is to find an identity of the type (\ref{identityMCG}) for a once punctured torus. Interestingly, the following identity is of the type (\ref{identityMCG}):
 \begin{equation}\label{identitytorus}
 \sum_{h\in \text{MCG}(\mathcal{R}_{1,1})}\frac{2~\text{sinc}^2\left(\tau_{h\cdot\gamma}\right)}{1+e^{l_{h\cdot \gamma}}}=1
 \end{equation}
 where, $\text{sinc}(x)=\frac{\text{sin}(\pi x)}{\pi x}$ is the normalized sinc function, and $\tau_{\gamma}$ is the Fenchel-Nielsen twist parameter defined with respect to the simple closed geodesic $\gamma$. Under the action of a full Dehn twist performed along $\gamma$ the twist parameter $\tau_{\gamma}$ becomes $\tau_{\gamma}+1$. Let us verify this identity. For this, note that a Dehn twist performed along the simple closed geodesic $\gamma$ has no effect on the geodesic length of $\gamma$. This means that we are allowed to do the following splitting 
  \begin{equation}
 \sum_{h\in \text{MCG}(\mathcal{R}_{1,1})}\frac{\text{sinc}^2\left(\tau_{h\cdot\gamma}\right)}{1+e^{l_{h\cdot \gamma}}}= \sum_{h_1\in \text{MCG}(\mathcal{R}_{1,1})/\text{Dehn}(\gamma)}\frac{1}{1+e^{l_{h_1\cdot \gamma}}}\sum_{h_2\in \text{Dehn}(\gamma)}\text{sinc}^2\left(\tau_{h_1h_2\cdot\gamma}\right)
 \end{equation}
 Now using the explicit action of Dehn twist on the Fenchel-Nielsen twist parameter we get the following result:
   \begin{equation}\label{identity}
\sum_{h_2\in \text{Dehn}(\gamma)}\text{sinc}^2\left(\tau_{h_1h_2\cdot\gamma}\right)=\sum_{k\in\mathbb{Z}}\text{sinc}^2\left(\tau_{h_1\cdot\gamma}+k\right)=1
 \end{equation}
  Consequently, the identity (\ref{identitytorus}) is correct only if the following identity is true:
    \begin{equation}\label{McShanetorus}
 \sum_{\gamma}\frac{2}{1+e^{l_{\gamma}}}=1
  \end{equation}
  where, the sum is over all the simple closed geodesics on $\mathcal{R}_{1,1}$. Here we used the fact that the action of all elements in the group $ \text{MCG}(\mathcal{R}_{1,1})/\text{Dehn}(\gamma)$  on the simple closed geodesic $\gamma$ produce all the simple closed geodesics on  $\mathcal{R}_{1,1}$. Interestingly, this is the famous McShane identity for  hyperbolic once-punctured tori \cite{McShane1}. \par
  
  Let us analyze the identity (\ref{identitytorus}) to see whether it has the required properties (\ref{GPCO1}). The following fact in hyperbolic geometry is vey useful for this. The length of a geodesic on a hyperbolic Riemann surface passing through a collar with core geodesic $\gamma$ having length $c_*$ and not homotopic to $\gamma$ is of the order $e^{\frac{1}{c_*}}$ \cite{Taniguchi}. This implies that the length $l_{g\cdot \gamma}$ for all $g\in \text{MCG}(\mathcal{R}_{1,1})/\text{Dehn}(\gamma)$ is of the order $e^{\frac{1}{c_*}}$. The reason is that for such $g$ the simple closed geodesic $g\cdot \gamma$ passes through the collar and it is not homotopic to the core geodesic $\gamma$ \cite{Hatcher}. Using these facts and the identity  one finds that
  \begin{align}
   \lim_{l_{\gamma}\to c_*}\frac{2~\text{sinc}^2\left(\tau_{g\cdot\gamma}\right)}{1+e^{l_{ g\cdot\gamma}}}=\mathcal{O}\left(e^{-1/c_*}\right)\qquad\qquad &\forall \quad g\in \text{MCG}(\mathcal{R}_{1,1})/\text{Dehn}(\gamma)\nonumber\\~\nonumber\\ 
   \lim_{l_{\gamma}\to c_*}\sum_{h\in \text{Dehn}(\gamma)}\frac{2~Q(c_*)~\text{sinc}^2\left(\tau_{h\cdot\gamma}\right)}{1+e^{l_{ h\cdot\gamma}}}&=1+\mathcal{O}\left(c_*^3\right)
     \end{align}
     with $$Q(c_*)=1+\frac{1}{2}c_*+\frac{1}{4}c_*^2$$ Therefore, by identifying $G(\gamma)$ in (\ref{identityMCG}) with $$\frac{2~Q(c_*)~\text{sinc}^2\left(\tau_{\gamma}\right)}{(1+e^{l_{ \gamma}})}$$ and the constant $Q$ in (\ref{identityMCG})  with $Q(c_*)$, we obtain the following PCO distribution that is compatible with the gluing of NS punctures up to $\mathcal{O}(c_*^2)$:
  \begin{equation}\label{PCODisttorus}
  \sum_{h\in \text{MCG}} \frac{Q(c_*)~\text{sinc}^2\left(\tau_{h\cdot\gamma}\right)}{(1+e^{l_{ h\cdot \gamma}})}\left\{\chi( p_1^{h\cdot \gamma})+\chi( p_2^{h\cdot \gamma})\right\}
  \end{equation}\par~\par
  
\noindent{\bf{\underline{Gluing of R punctures} :}} Consider the situation where two R punctures on a thrice-punctured sphere are glued via plumbing fixture to obtain a once-punctured torus. In this case the when the length of a simple closed geodesic on the once-punctured torus becomes 
$c_*$, the PCO distribution on it must match with that on the thrice-punctured sphere with one NS and two R punctures up to $\mathcal{O}(c_*^2)$. On such a thrice-punctured sphere, we don't have to insert any PCO, instead we must smear a PCO on a cycle on the plumbing collar homotopic to the core geodesic on the collar, while we glue two R punctures. Hence, the  PCO distribution (\ref{PCODisttorus}) must to be modified in order to make it compatible with the gluing of two R punctures.\par

 Let us modify the PCO distribution as follows. Use the  PCO distribution (\ref{PCODisttorus}) on  once-punctured hyperbolic tori with no simple closed geodesic having length less than $c_*(1+\delta)$, where $\delta$ is the same infinitesimal parameter introduced in the previous section  for modifying the local coordinate around the punctures. On a once-punctured hyperbolic torus with a simple closed geodesic having length  than $c_*(1+\delta)$ change the PCO distribution (\ref{PCODisttorus}) by replacing $\frac{1}{2}\left\{\chi( p_1^{h\cdot \gamma})+\chi( p_2^{h\cdot \gamma})\right\}$ by  a PCO smeared on a cycle on the plumbing collar homotopic to the core geodesic of the collar.

\subsection{PCO distribution on general surfaces}

Now we need to consider more general situations, where the string diagram has many handles and punctures. Interestingly, the story is similar for any hyperbolic string diagram.  In order to construct a consistent PCO distribution on elementary string diagrams, the generalized McShane identity satisfied by the lengths of simple closed geodesics on a hyperbolic Riemann surface is very useful \cite{Mirzakhani1}. This will be described in detail in the second part of this series \cite{Roji}.

\acknowledgments

We thank Seyed Faroogh Moosavian, Ashoke Sen,  Edward Witten, and Barton Zwiebach for the helpful discussions. Research at Perimeter Institute is supported by the Government of Canada through Industry Canada and by the Province of Ontario through the Ministry of Research \& Innovation.

\appendix

\section{Eisenstein Series}\label{Eisen}

 In this appendix, we briefly discuss the definition and some properties of the Eisenstein series following \cite{Kubota}. Consider a discrete subgroup $\Gamma$ of $PSL(2,\mathbb{R})$ acting on the upper half-plane $\mathbb{H}$. Let $\kappa_1,\kappa_2,\cdots,\kappa_h$ be the set of all cusps of $\Gamma$ that are not equivalent with respect to $\Gamma$.  Denote the stabilizer of $\kappa_i$ in $\Gamma$ by $\Gamma_i$:
 \begin{equation}
 \Gamma_i=\left\{\sigma\in \Gamma\quad |\quad \sigma \kappa_i=\kappa_i\right\}
 \end{equation}
 Consider the transformation $\sigma_i\in \text{SL}(2,\mathbb{R})$ which maps $\infty$ to $\kappa_i$: $$\sigma_i\infty =\kappa_i$$ The transformation $\sigma_i$ is chosen such that $\sigma_i^{-1}\Gamma_i\sigma_i$ is equal to the group $\Gamma_0$ of all matrices of the form    
 $\left(\begin{array}{cc}1 & m \\0 & 1\end{array}\right)$ with $m\in \mathbb{Z}$. Then, the {\it Eisenstein series} $E_i(z,s)$ for the cusp $\kappa_i$ is defined by 
 \begin{equation}
 E_i(z,s)=\sum_{\sigma\in \Gamma_i\backslash \Gamma} \left\{\text{Im}\left(\sigma_i^{-1}\sigma z\right)\right\}^s
 \end{equation}
where $s$ is a complex variable. Whenever the series converges uniformly, the Eisenstein series have the following properties:
\begin{itemize} 
\item $E_i(\sigma z,s)=E_i(z,s)$ for any $\sigma\in \Gamma$;
\item $DE_i=s(s-1)E_i$, where $D$ denotes the Laplacian of $\mathbb{H}$;
\item $E_i$ does not depend on the particular choice of a cusp $x_i$ among equivalent ones;
\item $E_i(z,s)$ converges absolutely, if $\text{Re}(s)>1$.
\end{itemize}

\par

\noindent{\bf{\underline{Fourier expansion at a cusp}~:}} The Fourier expansion of $E_{i}(z,s)$ at $\kappa_j$ is as follows
\begin{align}
E_i(\sigma_j z,s)&=\delta_{ij}\left(\text{Im}~z\right)^s+\phi_{ij}(s)\left(\text{Im}~z\right)^{1-s}\nonumber\\
&+\sum_m 2\pi^s|m|^{s-\frac{1}{2}}\Gamma(s)^{-1} \text{Re}(z)^{\frac{1}{2}}K_{s-\frac{1}{2}}\left(2\pi |m|\text{Re}(z)\right)\phi_{ij,m}(s) e^{2\pi \text{i}m~ \text{Re}(z)}
\end{align}
where $\phi_{ij,m}(s)$ denotes the following summation
\begin{align}
\phi_{ij,m}(s)=\sum_{c,d}\frac{1}{|c|^{2s}}e^{2\pi\text{i}md/c}\qquad\qquad &c>0 \qquad d~\text{mod}~c\nonumber\\& \left(\begin{array}{cc}* & * \\c & d\end{array}\right)\in \quad \sigma_i^{-1}\Gamma\sigma_j
\end{align}
and $\phi_{ij}(s)$ is given by
\begin{equation}
\phi_{ij}(s)=\pi^{\frac{1}{2}}\frac{\Gamma\left(s-\frac{1}{2}\right)}{\Gamma(s)}\phi_{ij,0}(s)
\end{equation}
The matrix $\phi_{ij}(s)$ is symmetric, $\phi_{ij}(s)=\phi_{ji}(s)$.


\end{document}